\documentclass[pra,twocolumn,a4paper,showpacs]{revtex4}
\usepackage{graphicx}
\usepackage{bm}
\newcommand{\nn}{\nonumber}

\begin{document}
\title{Instabilities of off-centered vortices in a Bose-Einstein condensate}
\date{\today}

%----------------
\author{Tomoya Isoshima}
\email{tomoya@focus.hut.fi}
\author{Jukka Huhtam\"{a}ki}
\author{Martti M. Salomaa}
\affiliation{
Materials Physics Laboratory,
Helsinki University of Technology, \\
P.~O.~Box 2200 (Technical Physics), FIN-02015 HUT, Finland }

\begin{abstract}
We study numerically the excitations of off-centered vortices
in a Bose-Einstein condensate.
The displacement of a single vortex and
the separation of a doubly quantized vortex are considered.
We find that the core-localized excitations of
the precessing vortices continue to feature the property
which implies that the vortices are unstable.
The core-localized, dipolar, and quadrupolar excitations are
found to deform continuously as functions of vortex displacements and
intervortex separation.
\end{abstract}

%The results are interpreted in terms of
%the precessional motion of the vortices.

% 03.75.Lm Tunneling, Josephson effect,
%     Bose-Einstein condensates in periodic potentials, 
%     solitons, vortices and topological excitations
% 03.75.Kk 
%     Dynamic properties of condensates;
%     collective and hydrodynamic excitations, superfluid flow
% 05.30.Jp Boson systems
% 67.40.Vs Vortices and turbulence

\maketitle

\section{Introduction}

Quantized vorticity which accompanies rotational motion
is a characteristic feature of superfluids.
Vorticity involves singular lines, vortex lines
which exist inside any closed path enclosing finite circulation.
The change in the phase of the order-parameter field
around the vortex line is quantized in integer units of $2\pi$.

In a Bose-Einstein condensate (BEC) of an atomic gas~\cite{pethick},
the first vortex was created by the JILA group~\cite{matthews}.
They used a two-component condensate of $^{87}$Rb atoms
and imprinted a phase winding of $2\pi$ onto one component.
The position of the vortex core was clearly seen and
also the precessional motion of the core was observed~\cite{matthews,anderson}.
In a one-component condensate, the ENS group~\cite{ens1st} created a vortex state
using a rotating trap which consists of an optical spoon and a magnetic trap.
Condensates containing up to 4 vortices were observed.
Their method can be understood in analogy with fluid motion
in a rotating vessel.
The idea of a rotating trap has been employed by many groups thereafter.

In addition to these singly quantized vortices,
multiply quantized vortices with phase windings $4\pi$ and $8\pi$ can also
be formed~\cite{topological},
using a topological method introduced by
Nakahara \textit{et al.}~\cite{nakahara}.
However, the experimentally recorded image of the cloud
of atoms does not show any indications for the 
splitting of multiply quantized vortices.
This surprising apparent stability contradicts with what has been
widely expected~\cite{pethick,pu}.
Therefore, it is urgent to understand the stability
of these off-centered vortices and the multiply quantized vortices.

The axisymmetric and rotationally symmetric (singly quantized) vortex states
have been studied and a core-localized excitation with negative excitation
energy is found~\cite{dodd,tomoyaaxisym1,tomoyaaxisym2,sami,sami:tpopov,svidzinsky}.
Concerning this excitation,
a relation to the instability of a vortex
state~\cite{tomoyaaxisym2,svidzinsky} (through the transfer of 
the population from the condensate mode to the core-localized excitation) and
also to the precession motion~\cite{linn,feder} of the vortex core
have been pointed out.
Nevertheless, the axisymmetric and rotationally symmetric systems only mimic
the experimental situations because
(a) the axisymmetric vortex state can become a vortexfree state
only through the off-centered vortex states.
(b) a system with a vortex under precessional motion is no
longer axisymmetric.

The understanding of the excitation level with negative energy
involves a puzzling feature~\cite{comparison}
because the value is connected to both
the direction and frequency of the precession motion
and the instability of the vortex state.
These two cannot be divided as long as we study the axisymmetric state.
Therefore, investigations of the off-centered vortex state are
necessary in order to understand both the instability and the precessional motion.
The off-centered vortex state has been studied
through the Gross-Pitaevskii equation~\cite{butts} and the
Thomas-Fermi (TF) approximation~\cite{svidzinsky,guilleumas}.
But no solutions of the Bogoliubov equations~\cite{pethick} have thus far been
presented for a non-axisymmetric vortex state.

The motivation for an analysis of non-axisymmetric doubly quantized vortex
may be explained in a similar way:
The instability of the doubly quantized vortex state has been pointed
out for the rotationally symmetric vortex state~\cite{pu}.
It relates the existence of a mode with a complex excitation energy 
and its exponential increase of population.
But the axisymmetric state only lasts for an infinitesimal time
if the vortex is unstable.
It is not known whether the system keeps having the
feature (complex eigenvalue) on which the instability of doubly 
quantized vortex state depends.

This paper presents a two-dimensional analysis of
the excitation spectra supported by
the condensate 
with an off-centered vortex, a doubly quantized vortex,
and pairs of split vortices.
We discuss the above questions
of precession, the instability of singly quantized vortex,
and the doubly quantized vortex within a unified numerical framework.

%=================================================================
\section{Single off-centered vortex}\label{sec:single}

We consider a two-dimensional $(x,y)$ system
in the rotating frame whose angular velocity is $\omega$.
The axis of rotation is perpendicular to the $(x,y)$ plane.
Ideally, this represents not only the static condensate
in the rotating vessel but also the precessing vortex.
The Hamiltonian is
\begin{eqnarray}
    \hat{H}({\bm{\omega}}) &=& \hat{H} - {\bm{\omega}}\cdot\hat{L}
\\
    \hat{H} &=&
        \int \hat{\Psi}^{\dagger} (-C \nabla^2 + V)\hat{\Psi}
	+ \frac{1}{2} g \hat{\Psi}^{\dagger} \hat{\Psi}^{\dagger}
	                \hat{\Psi}\hat{\Psi}
	d\mathbf{r}
\\
    \hat{L} &=&
        \int \hat{\Psi}^{\dagger} (\mathbf{r}\times\mathbf{p}) \hat{\Psi}
    d\mathbf{r}
\end{eqnarray}
where
$C = -\hbar^2/(2m)$, $g = 4\pi\hbar^2a/m$, and $V$ is the confining potential
with $a$ denoting the s-wave scattering length and $m$ the mass.
We use ${\bm{\omega}} = (0,0,\omega)$ and $\mathbf{r} = (x,y,0)$.
The time-dependent Gross-Pitaevskii (GP) equation may be written as
\begin{eqnarray}
\{
    - C \nabla^2 + V(x,y) + g|\phi(x,y,t)|^2 &&\nn
\\
    - {\bm{\omega}}\cdot\mathbf{r}\times \mathbf{p}
\} \phi(x,y,t) &=& i \hbar \frac{\partial \phi(x,y,t)}{\partial t},
\label{eq:gpt}
\end{eqnarray}
where $\phi(\mathbf{r})$ is the condensate wavefunction.
The time-independent form is
\begin{eqnarray}
\{
    - C \nabla^2 + V(x,y) - \mu + g|\phi(x,y)|^2 &&\nn
\\
    - {\bm{\omega}}\cdot\mathbf{r}\times \mathbf{p}
\} \phi(x,y) &=& 0,
\label{eq:gp}
\end{eqnarray}
where $\mu$ is the chemical potential.
We employ an axisymmetric harmonic trapping potential
$V(x,y) = m\omega_{\mathrm{tr}}^{2}({x^2 + y^2})$
where $\omega_{\mathrm{tr}} = 2\pi \times 200$.
The linear number density of particles
$N = \int |\phi(\mathbf{r})|^2 d\mathbf{r}$
is fixed to $1\times 10^{10} m^{-1}$ in this paper.
The mass $m = 38.17 \times 10^{-27} \text{ kg}$ and
the scattering length $a = 2.75 \text{ nm}$ for Na atoms 
are employed.

The angular momentum of the condensate per particle
\begin{equation}
    L = \frac{
        \int \phi^{\ast}(\mathbf{r} \times \mathbf{p})\phi \, d\mathbf{r}
	}{ \hbar \int |\phi(\mathbf{r})|^2 d\mathbf{r} }
   \label{eq:L}
\end{equation}
is extensively utilized in the following discussion
and plotted in several figures.
A condensate with a centered vortex has $L = 1$.
Systems with an off-centered vortex have $L$ between 0 and 1.
We also use the TF radius 
$R_{\rm{TF}} = 6.79 \, \mu {\rm m}$ as the unit of length.

%================================================================
%\subsection{Condensate Wavefunctions}
\subsection{Static Analysis}\label{time-dependent}

The time-dependent Bogoliubov equations~\cite{sami:tpopov}
\begin{eqnarray}
\lefteqn{
i\hbar \frac{\partial u(x,y,t)}{\partial t}
}\nn\\
&=&
(
    -C\nabla^2 + V - \mu + 2g|\phi|^2
    -{\bm{\omega}}\cdot\mathbf{r}\times \mathbf{p}
)u(x,y,t)
\nn\\*
   &&{} - g\phi^2 v(x,y,t), 
   \label{eq:tbog1}
\\
\lefteqn{
i\hbar \frac{\partial v(x,y,t)}{\partial t}
}\nn\\
&=&
    -\left(-C \nabla^2 + V - \mu + 2g|\phi|^2
        + {\bm{\omega}} \cdot \mathbf{r} \times \mathbf{p}
    \right)v(x,y,t)
\nn\\*
   && {} + g\phi^{\ast 2} u(x,y,t),
   \label{eq:tbog2}
\end{eqnarray}
yield the excitation spectra of the condensate which follows the
time-dependent GP equation Eq.~(\ref{eq:gpt}).
They reduce to the Bogoliubov equations~\cite{pethick}
\begin{eqnarray}
    \samepage
    \left(-C \nabla^2 + V - \mu + 2g|\phi|^2
        - {\bm{\omega}} \cdot \mathbf{r} \times \mathbf{p}
    \right)u
\nn\\*
    - g\phi^2 v &=& \varepsilon u , 
    \label{eq:bog1}
\\
    -\left(-C \nabla^2 + V - \mu + 2g|\phi|^2
        + {\bm{\omega}} \cdot \mathbf{r} \times \mathbf{p}
    \right)v
\nn\\*
    + g\phi^{\ast 2} u &=& \varepsilon v.
    \label{eq:bog2}
\end{eqnarray}
when the system is static.

The criteria for the applicability 
of the Bogoliubov Eqs.~(\ref{eq:bog1}) and (\ref{eq:bog2}),
instead of the time-dependent
Bogoliubov Eqs.~(\ref{eq:tbog1}) and (\ref{eq:tbog2}),
depends on how fast the wavefunction of the condensate changes.
We measure the amplitude of the transformation using
\begin{equation}
   V^{\prime} \equiv \frac{
   \max\left(\left|
       ( - C \nabla^2 + V + g|\phi|^2
       - {\bm{\omega}}\cdot\mathbf{r}\times \mathbf{p}- \mu)\phi
   \right|\right)
   }{ \hbar \cdot \max(|\phi|) }
    \label{eq:velocity}
\end{equation}
where the chemical potential $\mu$ is fixed here~\cite{chemical1,chemical2}.
The numerator is a maximum of the left-hand side of Eqs.~(\ref{eq:gpt}) and (\ref{eq:gp}).
Therefore, in the framework of the time-dependent GP equation
\begin{equation}
    V^{\prime}(t) = \max\left(\left| \frac{\partial\phi(x,y,t)}{\partial t} \right|\right) /
    \max(|\phi(x,y,t)|).
    \label{eq:velocity2}
\end{equation}
For variations of the displacement of the vortex core $r$,
the $V^{\prime}$ has local minima
for an axisymmetric vortex state, a vortexfree state,
and an off-centered vortex state with certain displacement 
$r$ for a choice of $\omega$.

We use off-centered vortex states which satisfy
\begin{equation}
    V^{\prime} < 0.0005 \label{eq:condition}
\end{equation}
in the following analysis of the Bogoliubov equations.
Equations (\ref{eq:velocity2}) and (\ref{eq:condition}) imply that
the wavefunction $\phi$ varies slowly enough
in the time scale of the harmonic trap
[$V^{\prime} < 0.0005 \ll \omega_{\mathrm{tr}}/(2\pi)$].
Instead of the right-hand side of Eq.~(\ref{eq:condition}),
for low values of the angular momentum ($L < 0.2$)
the right-hand side is rather relaxed to $0.03$.

\begin{figure}
\begin{center}
\includegraphics[width=6.5cm]{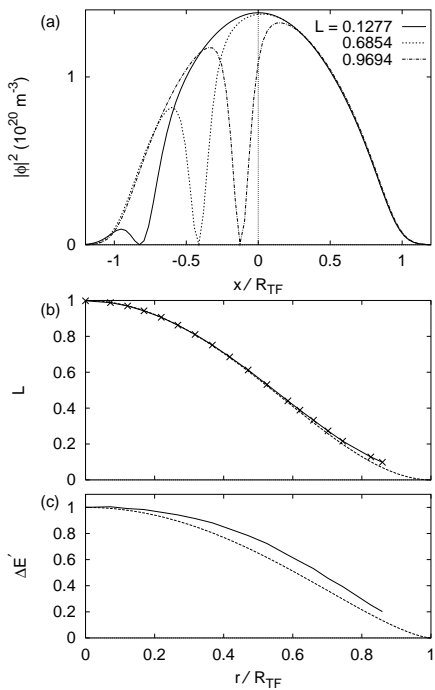}
\end{center}
\caption{\label{fig:dns1}
(a) Density of the condensate along the $x$-axis.
The angular momenta $L$ are 0.1277, 0.6854, and 0.9694, respectively.
(b) The solid line represents the
displacement of a vortex core vs. angular momentum of the condensate per particle.
The displacement vanishes for $L=1$.
The dotted line is the analytical estimate $\{1 - (r/R_{\mathrm{TF}})^2\}^{2}$
by Guilleumas~\cite{guilleumas}.
These two plots overlap in a wide range of
displacements, $0 \le r < 0.8$.
(c) Displacement of the vortex core $r$ vs.
the normalized additional energy $\Delta E^{\,\prime}$ of the condensate
in the presence of an off-centered vortex.
The solid line represents the numerical result.
The dotted line is obtained by fitting $\{1 - (r/R_{\mathrm{TF}})^2\}^{3/2}$,
taken from Ref.~\cite{fetter:vortex}, Eq.~(49).
}
\end{figure}

Each of the relaxation processes,
which is numerically equivalent to following the
time-dependent GP equation with imaginary time,
begins with the TF density profile as the initial configuration
on which the phase of an off-centered vortex
with various displacements $r$ is ``imprinted''
through multiplication.
In the beginning of a numerical simulation,
the angular velocity $\omega$ is chosen.
Then the development with fixed $\omega$ is performed
without restricting the displacement.
The vortex moves freely during the development and
ends as an axisymmetric vortex,
a vortexfree state,
or a vortex state which satisfies Eq.~(\ref{eq:condition}).
The latter form a set of $\phi$ and $\omega$ values
which fulfill Eq.~(\ref{eq:condition}).
Figure \ref{fig:dns1}(a) shows some of the density profiles $|\phi|^2$ 
of the condensate.
The displacement of the vortex core $r$ and 
the angular momentum $L$ are calculated through Eq.~(\ref{eq:L}) from each of
the resulting wavefunctions $\phi$.
Therefore, $L$ and $r$ are functions of $\omega$ through $\phi$.
The relation between the angular momentum and the displacement is
plotted in Fig.~\ref{fig:dns1}(b).

% The angular velocities $\omega$ that have been determined in this way
% are discussed in Sec.~\ref{sec:precession}.

Figure \ref{fig:dns1}(c) displays the normalized additional energy
due to the existence of an off-centered vortex, defined through
\begin{equation}
    \Delta E^{\prime} = (E - E_{0} ) / (E_{1} - E_{0})
\end{equation}
where $E_0$, $E_1$, and $E$ are
the energies of a vortexfree condensate,
a condensate in the presence of a centered vortex,
and with an off-centered vortex, respectively.
If the condensate is vortexfree, $\Delta E^{\prime} = 0$.
Off-centered vortex states have $0 < \Delta E^{\prime}< 1$
and the centered vortex state has $\Delta E^{\prime} = 1$.

%------------------------------
%------------------------------
%------------------------------
\begin{figure}
\begin{center}
\includegraphics[width=6cm]{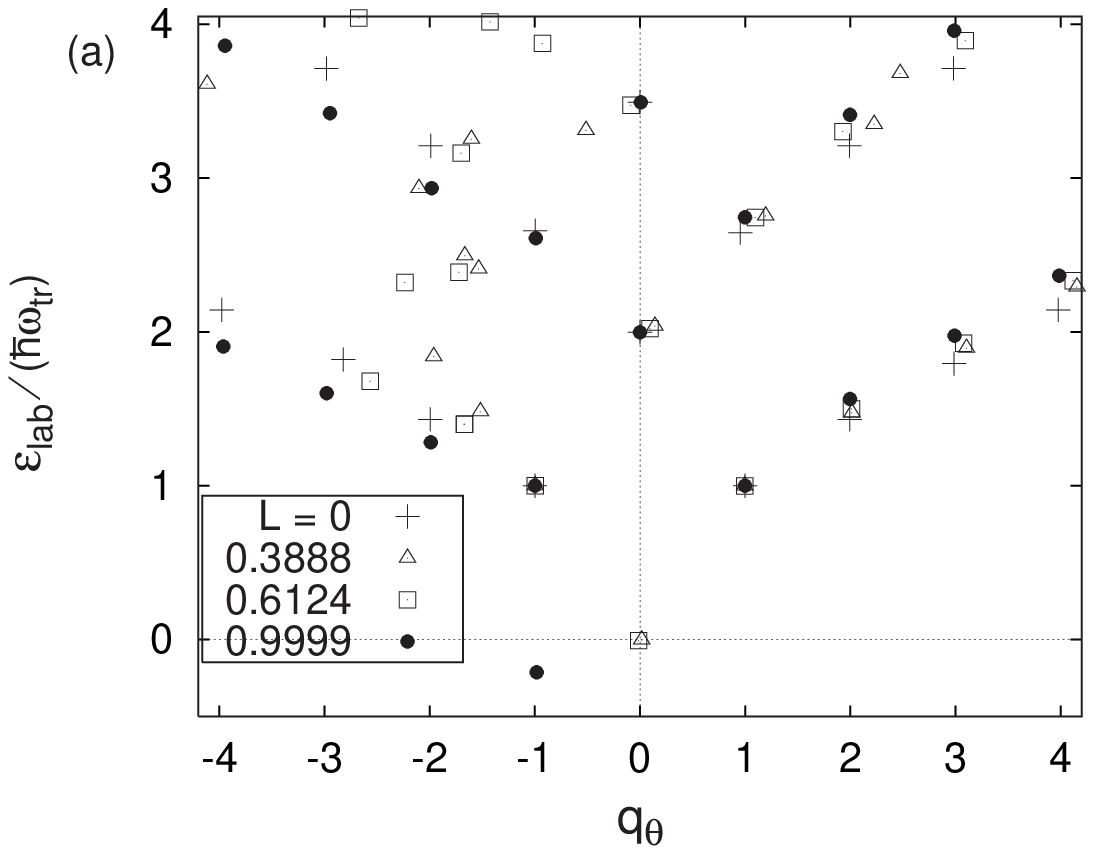}\\
\includegraphics[width=6cm]{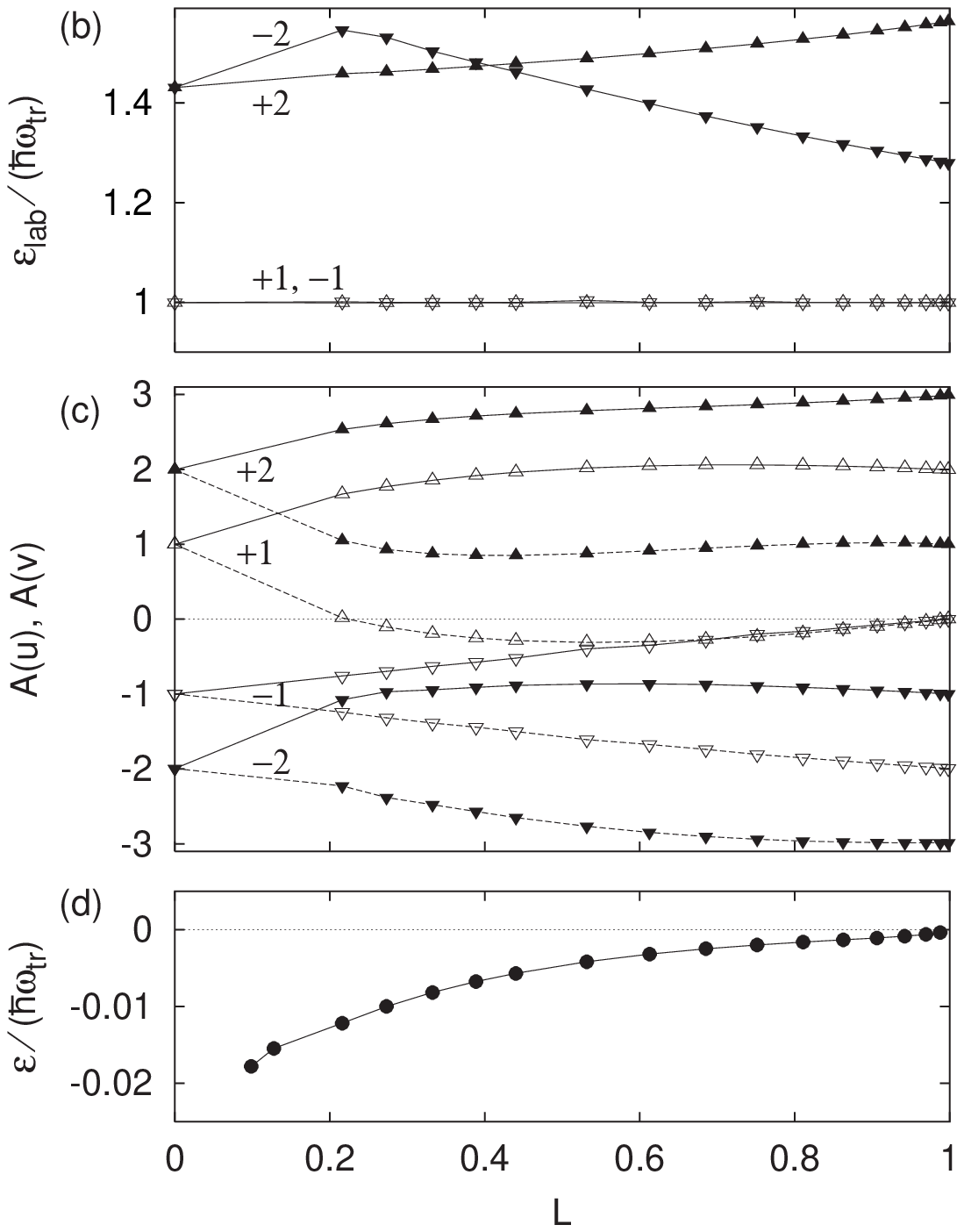}
\end{center}
\caption{\label{fig:di_quadru_1}
(a) Excitation spectra of a Bose-Einstein condensate with a centered vortex,
an off-centered vortex, and without a vortex.
The vertical scale measures the excitation energy, $\varepsilon_{\mathrm{lab}}$.
The horizontal axis indicates the angular momentum $q_{\theta}$.
The crosses, triangles, squares, and the bullets correspond to 
the angular momenta of the condensate of $L = 0, 0.38, 0.61, \text{ and } 1$, respectively.
(b) Excitation energies of the dipole modes (labelled $+1$ and $-1$) and
the quadrupole modes ($+2$, $-2$).
(c) The angular momenta of the components
$u$ (solid line) and $v$ (dotted line) of wavefunction for 
the dipole ($+1$, $-1$) and quadrupole modes ($+2$, $-2$).
(d) Energy $\varepsilon$ of the lowest core excitation in the rotating frame
is negative and it satisfies $|\varepsilon| \ll 1$.
This is in accordance with the weak instability of the GP equation, Eq.~(\ref{eq:condition}).
}
\end{figure}
%-----------------------------/

%----------------------------------------------------------------
\subsection{Excitation Spectra}

Excitations from a BEC have been interpreted in the framework of
the Bogoliubov equations in Ref.~\cite{vogels}.
They play a significant role in determining vortex stability. 
Nevertheless, these modes in off-centered vortices have not been analyzed so far.
An excitation whose energy is $\varepsilon$
is expressed with two components, $u$ and $v$, of wavefunction each with different momenta.
They are determined as the solutions to
the eigensystem of Bogoliubov Eqs.~(\ref{eq:bog1}) and (\ref{eq:bog2}).
The modes with negative $\int (|u|^2 - |v|^2) d\mathbf{r}$ are ignored.
Two of the lowest excitations are
the condensate mode (with $\varepsilon = 0, u = \phi, v = \phi^{\ast}$) and
the core-localized mode.
We ignore the mode with the smallest $|\varepsilon|$ 
as the condensate mode.

The wavefunction expressed with $u$ and $v$ are obtained through 
the Bogoliubov equations (\ref{eq:bog1}) and (\ref{eq:bog2}).
We denote their angular momenta using
\begin{eqnarray}
    \mathcal{A}(u) &\equiv&
    \frac{
        \int u^{\ast}(\mathbf{r} \times \mathbf{p})u \, d\mathbf{r}
    }{
        \hbar \int |u|^2 d\mathbf{r}
    },
\\
    \mathcal{A}(v) &\equiv&
    \frac{
        \int v^{\ast}(\mathbf{r} \times \mathbf{p})v \, d\mathbf{r}
    }{
        \hbar \int |v|^2 d\mathbf{r}
    }.
\end{eqnarray}
These quantities correspond to
the angular momentum of the condensate $L$ defined in Eq.~(\ref{eq:L}).
Some algebra with the help of Eqs. (\ref{eq:gp}), (\ref{eq:bog1}), and (\ref{eq:bog2})
shows that the excitation energies $\varepsilon$ depend linearly on $\omega$.
We presume that the wavefunctions $\phi$, $u$, and $v$ do
not depend on $\omega$ and energies $\mu$ and $\varepsilon$ do not have imaginary part.
The coefficient of proportionality is
\begin{equation}
    q_\theta \equiv 
    \frac{
        \mathrm{Re}\left[
            \{ \mathcal{A}(u) - L \} \int |u|^2 d\mathbf{r} +
            \{ \mathcal{A}(v) + L \} \int |v|^2 d\mathbf{r}
        \right]
    }{
        \int (|u|^2 + |v|^2) d\mathbf{r}
    } . \label{eq:q_theta} 
\end{equation}
We introduce an excitation energy in the laboratory frame
\begin{equation}
    \varepsilon_{\mathrm{lab}} \equiv \varepsilon + \hbar \omega q_{\theta}.
    \label{eq:e_lab}
\end{equation}
Equation (\ref{eq:q_theta}) includes $\mathcal{A}(u)$, $\mathcal{A}(v)$,
and $L$ because an excitation level has two components $u$ and $v$ of wavefunction
and their phases are always affected by the phase of the condensate.
For axisymmetric systems,
the $q_\theta$ definition reduces to
an integer quantum number~\cite{tomoyaaxisym1,tomoyaaxisym2}.
The excitation energies $\varepsilon$ and $\varepsilon_\mathrm{lab}$ 
are normalized by the trap unit $\hbar\omega_{\mathrm{tr}}$ from here on.
Figure \ref{fig:di_quadru_1}(a) displays the computed excitation energies,
$\varepsilon$, and
the corresponding angular momenta, $q_{\theta}$.
Some of the modes, for example those with
$(\varepsilon_\mathrm{lab}, q_{\theta}) = (2,0), (1,\pm 1)$
maintain similar values of
$q_\theta$ and $\varepsilon_\mathrm{lab}$ upon variations of $L$.
We call the modes at $(2,0)$ and $(1, \pm1)$
the breathing~\cite{pethick} and the dipole modes, respectively. 
Some of the core-localized modes have $(\varepsilon_{\mathrm{lab}}, q_{\theta}) = (0,0)$.
These are discussed in Sec.~\ref{sec:coremode}.
The modes with $(\varepsilon_{\mathrm{lab}}, q_{\theta}) \simeq (1,\pm 1), (1.5, \pm 2)$ are
discussed in Sec.~\ref{subsec:di_quadru}.

%------------------------------
\begin{figure}
\begin{center}
\includegraphics[width=8cm]{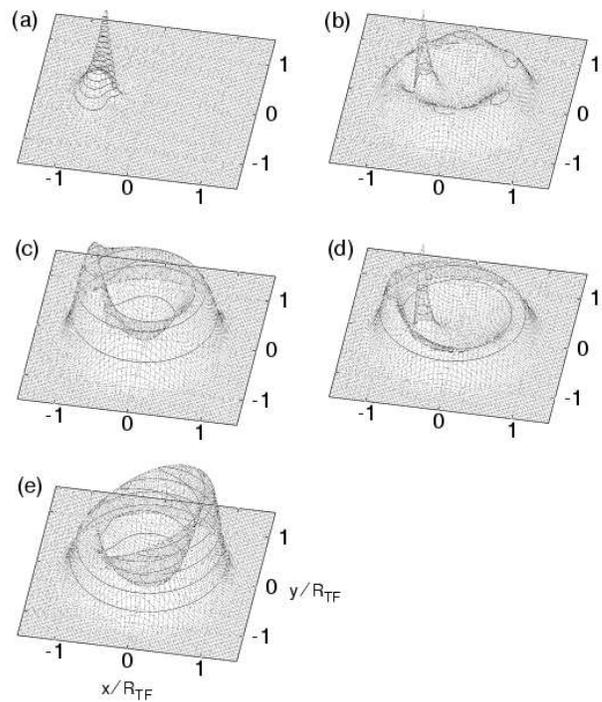}
\end{center}
\caption{\label{fig:wf_u}
Wavefunctions $|u|^2$ of (a) the core-localized mode,
(b) dipole mode with negative $q_\theta$,
(c) dipole mode with positive $q_\theta$,
(d) quadrupole mode with negative $q_\theta$, and
(e) quadrupole mode with positive $q_\theta$.
Solid lines denote contours of density.
The core-localized mode has a peak of the amplitude in the vortex core.
%The dotted line shows $R_{\mathrm{TF}}$.
The peak in the quadrupole mode with negative $q_\theta$ 
is not seen in axisymmetric systems.
The angular momentum $L$ of the condensate is 0.531.
}
\end{figure}
%------------------------------

%-------------------------------------------------------------------
\subsection{Core-Localized Mode}\label{sec:coremode}

For any value of $L$, the system has one core-localized mode.
The modes always displays a sharp peak in the amplitude $|u|^2 + |v|^2$ of 
the wavefunction at the vortex core.
The core-localized modes still feature a major difference between
the system with $L=1$ and those with $0<L<1$.

When $L = 1$, the vortex is in the center of the system.
The angular momentum $q_\theta = -1$ and 
the excitation energy $\varepsilon_{\mathrm{lab}}$ is negative
[the lowest bullet at $q_\theta = -1$ in Fig.~\ref{fig:di_quadru_1}(a)].
It has a sharp peak in the vortex core.
The sharp localization, the angular momentum, and the negative energy
are equivalent with those for
axisymmetric calculations~\cite{tomoyaaxisym1,tomoyaaxisym2}.

Once the vortex becomes off-centered ($0<L<1$),
both of the components $u$ and $v$ of the wavefunction localize in the vortex core
[Fig.~\ref{fig:wf_u}(a)].
Each of them features a sharp peak in the vortex core and
its skirt extends towards the surface of the condensate.
The angular momenta $\mathcal{A}(u)$ of the core-localized modes vary
between 1 and 4.
The corresponding wavefunctions $v$ have opposite angular momenta.
Their average $q_{\theta}$ almost vanishes ($|q_\theta| \ll 1$);
the points at (0,0) in Fig.~\ref{fig:di_quadru_1}(a) mean this.
The excitation energy plotted in Fig.~\ref{fig:di_quadru_1}(d) remains 
below zero ($-0.01 < \varepsilon < 0$) even in the rotating frame.

The core-localized mode behaves differently
between the system with $L=1$ and that with $0<L<1$.
One possible cause is
the static approximation of the time-dependent Bogoliubov equations
introduced in Sec.~\ref{time-dependent}.
While the system with $L=1$ is treated with the Bogoliubov equations,
systems with $0<L<1$ are described within
the time-dependent Bogoliubov equations.

%----------------------------------------------------------------
\subsection{Dipole and Quadrupole Modes}\label{subsec:di_quadru}

Excitations having the lowest positive energy 
at $q_\theta = \pm 1$ and $\simeq \pm 2$ are
classified as the dipole and quadrupole modes.
Figure \ref{fig:di_quadru_1}(b) shows the energy for these modes.
The energies of the dipole modes almost equal
one trap unit $(\hbar\omega_{\mathrm{tr}})$.
The dipole modes are known to be responsible for the center-of-mass
motion~\cite{zambelli}.
Therefore, they are not affected by various profiles inside the condensate.
The angular momenta $\mathcal{A}(u)$ and $\mathcal{A}(v)$
are plotted in Fig.~\ref{fig:di_quadru_1}(c).
These excitations show continual transformation 
as a function of variation of the vortex displacement.

The condensate in a two-dimensional harmonic potential
has two quadrupole modes.
The splitting between the quadrupole frequencies
$\varepsilon_{\mathrm{lab}}(+2) - \varepsilon_{\mathrm{lab}}(-2)$
is inverted at $L \le 0.4$,
while it is proportional to $L$ in systems with
a centered vortex~\cite{zambelli}.
The corresponding eigenoscillation of the condensate, called the scissors mode,
has been observed experimentally~\cite{chevy}
to measure an angular momentum of the condensate.
The inversion may affect the observations of the angular momentum.

Guilleumas \textit{et al.} \cite{guilleumas} estimate
the splitting using several methods within the TF limit.
Their Eqs. (33) and (42) and our numerical results
in Fig.~2(b) agree well when the vortex is close to the axis
($r / R_{\mathrm{TF}} < 0.2$).
Their estimates stay positive and approach zero as the displacement of the vortex
approaches $R_{\mathrm{TF}}$.
But the sign is inverted in our numerical results.
This is caused by an increase of the energy $\varepsilon_{\mathrm{lab}}(-2)$
at lower $L$, depicted in Fig.~\ref{fig:di_quadru_1}(b).
The wavefunction $u$ of the $-2$ mode displays
a sharp peak at the vortex center;
for example, in Fig.~\ref{fig:wf_u}(d).
This peak is significant for lower $L$ and
not seen in the axisymmetric (non)vortex states.
We presume that the existence of this peak
is pushing up the energy $\varepsilon_{\mathrm{lab}}(-2)$.

For $\omega \gtrsim 0.35 \omega_{\mathrm{tr}}$ (corresponding to $L < 0.2$),
the excitation energies of the modes with $q_{\theta} > 2$ approach
each other,
and their wavefunctions become numerically indistinguishable 
from one another.
The angular momenta of the excitations in this region ($L < 0.2$)
are omitted in Figs.~\ref{fig:di_quadru_1} for the sake of clarity.

%=================================================================

\section{Two vortices}\label{sec:double}

A vortex with the winding number two has been formed in a condensate of
Na atoms~\cite{topological}.
Multiply quantized vortices are energetically unfavorable
against the array of the singly quantized vortices 
(e.g. Sec.~9.2.1 in Ref.~\cite{pethick}),
while the experimental data does not indicate any signs of the splitting.
In this section,
we investigate the excitation spectra of systems with two vortices
to inspect their stability in the splitting process.

\begin{figure}
\begin{center}
\includegraphics[width=6cm]{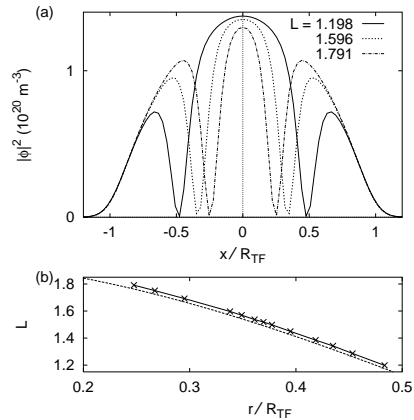}
\end{center}
\caption{\label{fig:dns2}
(a) Density of condensate along $x$-axis.
A condensate has two vortices and both of them are on the $x$-axis.
Angular momenta $L$ are 1.198, 1.596, and 1.791, respectively.
(b) Displacement of vortex core vs. angular momentum of the condensate par particle.
The displacement will be zero when the angular momentum of the condensate is 2.
The dotted line is the analytical estimate $2\{1 - (r/R_{\mathrm{TF}})^2\}^{2}$
taken from Eq.~(24) in Ref.~\cite{guilleumas}.
The two plots agree well.
}
\end{figure}

Each of the GP calculations starts from an initial wavefunction 
with two slightly displaced vortices under a fixed angular velocity $\omega$,
which acts as an external adjustable parameter.
The initial wavefunction is formed from the TF density profile
in the absence of a vortex, but multiplied
with the phase factor $e^{i\theta}$ around each of the initial vortex positions.
The resulting angular momenta are controlled through this $\omega$.
The GP equation has a static solution for about $60 < \omega/(2\pi) < 110$.
The corresponding range of angular momenta is from 1.2 to 1.8.
Many vortices enter into the condensate for higher $\omega/(2\pi) > 110$ and
the vortices disappear for $\omega/(2\pi) < 60$.
Figure \ref{fig:dns2} shows density profiles of the condensate for
various angular momenta.

%------------------------------
%------------------------------
%------------------------------
\begin{figure}
\begin{center}
\includegraphics[width=6cm]{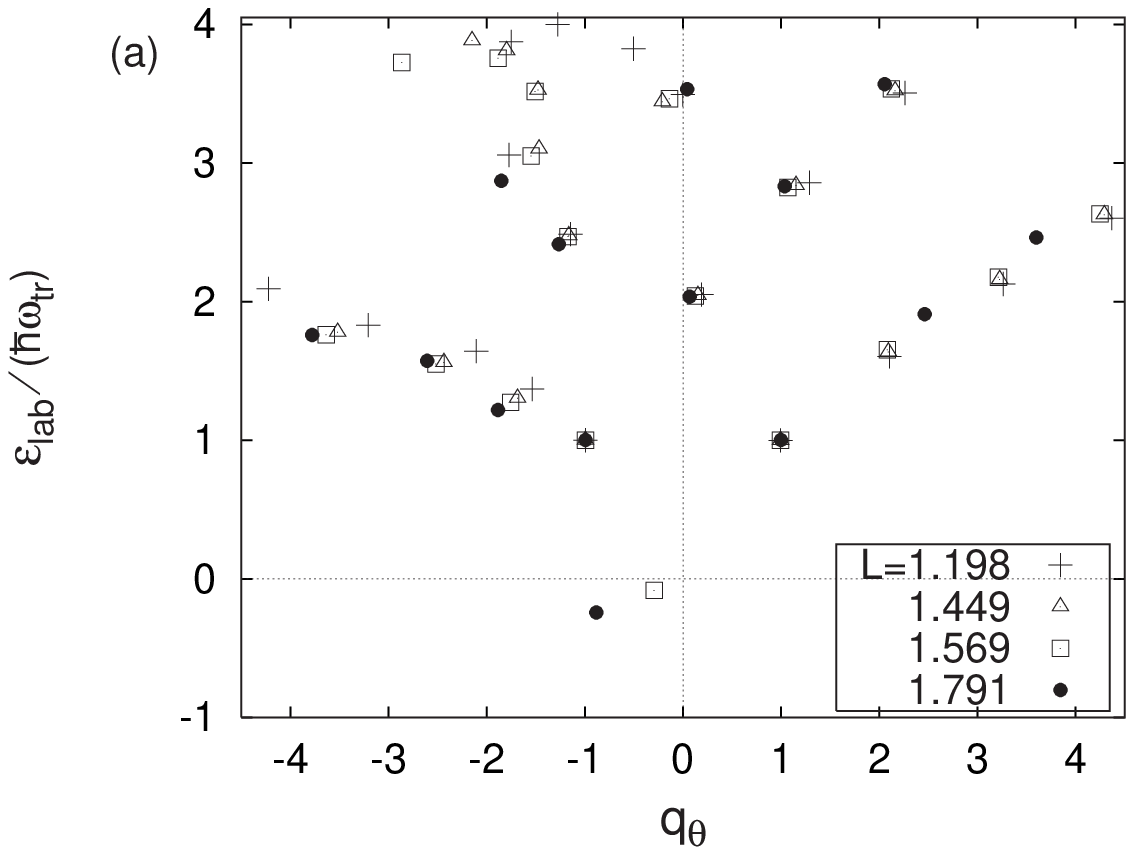}\\
\includegraphics[width=6cm]{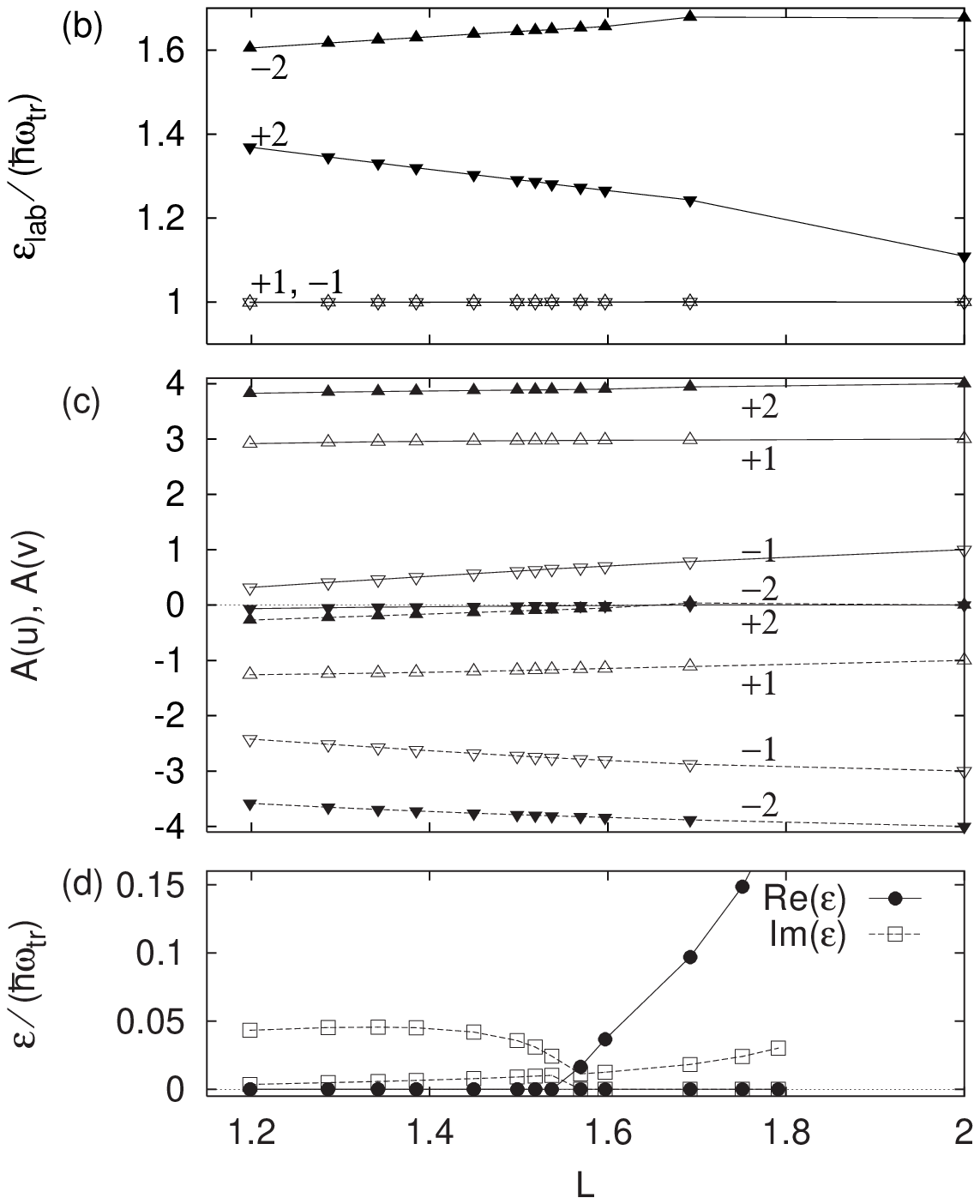}
\end{center}
\caption{\label{fig:di_quadru_2}
(a)
Excitation spectra of a Bose-Einstein condensate with a pair of
off-centered vortices.
The vertical axis denotes the excitation energy, $\varepsilon_{\mathrm{lab}}$.
The modes with complex energies are not plotted here
because the definitions of $\varepsilon_{\mathrm{lab}}$ and $q_{\theta}$
in Eqs.~(\ref{eq:q_theta}) and (\ref{eq:e_lab}) do not support
modes with complex energies.
The horizontal axis is the angular momentum, $q_{\theta}$.
The crosses, triangles, squares, and bullets correspond to 
the angular momenta of the condensate 1.198, 1.449, 1.569, and 1.791, respectively.
(b) Energy levels $\varepsilon_{\mathrm{lab}}$ of
the quadrupole and dipole modes in the laboratory frame.
(c) Angular momenta of the quadrupole and dipole modes.
Solid lines show those of $u$, while the dashed lines correspond to $v$.
(d) Two excitations with the lowest Re($\varepsilon$).
Conjugate modes with negative imaginary parts are not plotted.
A system has two conjugate pairs of complex eigenvalues or
one pair of complex eigenvalues and one real eigenvalue.
The real and imaginary parts are plotted independently.
The points at $L=2$ in (b) and (c) are taken from axisymmetric calculations.
}
\end{figure}
%-----------------------------/
%%%%%%% breathing modes 
%------------------------------
\begin{figure}[hbt]
\begin{center}
\includegraphics[width=5cm]{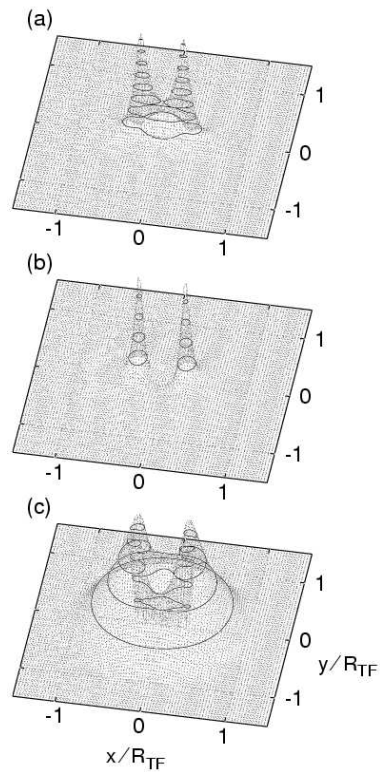}
\end{center}
\caption{\label{fig:core2}
(a) Wavefunction $|u|^2$ of a complex mode;
the wavefunction $v$ of the complex mode is the conjugate of $u$.
The energy levels of these modes are $\pm 0.0241i$.
(b) (c) Wavefunctions $|u|^2$ and $|v|^2$ of the lowest mode.
% The dotted line shows $R_{\mathrm{TF}}$.
The energy eigenvalue of this mode is $\varepsilon_{\mathrm{lab}} = -0.2329$.
The angular momentum of the condensate is 1.750.
}
\end{figure}
%------------------------------

Using the condensate wavefunctions,
the excitations from it are calculated with 
the Bogoliubov equations, Eqs.~(\ref{eq:bog1}) and (\ref{eq:bog2}).
We found that
one or two conjugate pairs of modes with complex eigenvalues 
always exists in the calculated range of $L$.

Figure \ref{fig:di_quadru_2}(d) shows the energy levels for
two excitations with the lowest real part in the rotating frame.
For $1.2 < L < 1.57$, there occur two conjugate pairs of complex eigenvalues,
both of them have a large imaginary part.
For larger $L$, only one of them has a large imaginary part.
Figure \ref{fig:core2} displays the wavefunctions of these lowest modes.
One of them has an imaginary eigenvalue,
while the other one has a real eigenvalue.
Both of them display a peak in the vortex core.

A doubly quantized vortex in a rotationally symmetric system
has two negative excitations, those for $q_\theta = -1$ and $-2$.
One of them with $q_\theta = -2$ sometimes has a complex energy,
depending on the particle density in the system, see Ref.~\cite{pu}.
This feature is equivalent to those of our results
depicted in Fig.~\ref{fig:di_quadru_2}(d) for higher $L$.
Therefore, the splitted vortices keep on having unstable nature
which exist before the splitting~\cite{pu}.

% The pairs of complex modes for lower $L$ in Fig.~\ref{fig:di_quadru_2}(d)
% may be understood as an entanglement of
% the complex core-localized modes for each of the singly quantized vortices
% discussed in Sec.~\ref{sec:coremode}.

Figure \ref{fig:di_quadru_2}(a) shows that
the modes with positive energy 
stay with similar values of $\varepsilon_\mathrm{lab}$ and $q_\theta$
as functions of the separation of the vortices.
The lowest positive-energy modes,
with $q_\theta \simeq \pm1$ and $q_\theta \simeq \pm2$,
are classified as the dipole and quadrupole modes.
Their computed angular momenta and energies in the
laboratory frame are presented in 
Fig.~\ref{fig:di_quadru_2}(b) and \ref{fig:di_quadru_2}(c).
The modes near $(q_\theta, \varepsilon_{\mathrm{lab}}) = (0,2)$
are breathing modes.
The consistent behavior of the dipole, quadrupole, and breathing modes
throughout the results in
systems with an off-centered vortex 
[Figs.~\ref{fig:di_quadru_1}(a)-\ref{fig:di_quadru_1}(c)],
two vortices
[Figs.~\ref{fig:di_quadru_2}(a)-\ref{fig:di_quadru_2}(c)],
and an axisymmetric vortex ($L =1, 2$)
proves the validity of our numerical procedure.

% Once the excitation energy becomes complex valued, 
% the distinguishness between the corresponding wavefunction $u$ and $v^\ast$
% disappears and the value $q_\theta$ in Eq.~(\ref{eq:q_theta}) becomes 0.
% The definition of $\varepsilon_{\mathrm{lab}}$ loses sence.
% Therefore the complex modes are not plotted in Fig.~\ref{fig:di_quadru_2}(a).

%-------------------------
\begin{figure}[thb]
\begin{center}
\includegraphics[width=7cm]{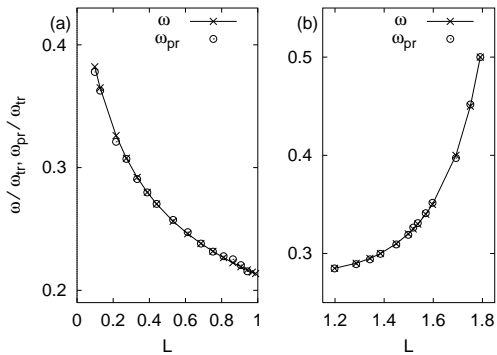}
\end{center}
\caption{\label{fig:precession}
Circles represent the precessional angular velocities 
according to time-dependent calculations;
lines show the angular velocity $\omega / \omega_{\mathrm{tr}}$ that
gives each value of angular momentum $L$.
The vertical axis is normalized by the trap frequency.
(a) Systems with one vortex. (b) Two vortices.
}
\end{figure}
%-------------------------

%----------------------------------------------------------------
\section{Precession Frequencies}\label{sec:precession}

The above results are calculated in the rotating frame.
It is not self-evident whether 
they may be interpreted as results for precessing
vortices in the laboratory frame.
To estimate the validity of the condensate wavefunction,
we carry out time-dependent Gross-Pitaevskii calculations in the laboratory frame.
The above results of the static GP equation are allowed to 
evolve in time according to Eq.~(\ref{eq:gpt}).

The condensate shows precessional motion of the vortices
for both the single-vortex and the two-vortex cases.
Figure \ref{fig:precession} shows
the precessional angular velocity, $\omega_{\mathrm{pr}}$.
The rotation frequency $\omega$ that we obtained in
the static calculations in
Secs.~\ref{sec:single} and \ref{sec:double} are also plotted.
The two frequencies agree well.
It means that our results in the rotating frame can also be identified as
those in the laboratory frame.
The excitation modes, including the core mode,
precess following the vortex core there.

%----------------------------------------------------------------
\section{Discussion}

The excitation spectra of Bose Einstein condensates with
one and two off-centered vortices has been computed.
The excitation with negative energy and localization at the vortex core
always exists not only in the axisymmetric state
but also for the off-centered vortex configurations.
We depicted the shape of wavefunctions and excitation energy
of the core-localized, dipole and quadrupole modes in
the off-centered vortex states with both one and two vortices.
The invariance of dipole and breathing modes
in the nonaxisymmetric cases shows
that usage of of the Bogoliubov equation instead of the time-dependent
Bogoliubov equation is valid.

The rotation frequencies agree well with the precession frequencies
of vortices in the laboratory frame.
It enables us to consider the two problems, precession of the vortices
and the vortex instability concerning the negative excitation, independently.
The precessing vortices pursue to have
a core-localized excitation with negative excitation energy
in the framework of Bogoliubov equations.
Therefore, the condensate with one precessing vortex
maintains the instability as long as the vortex exists.
In order to pose an answer to 
the problem~\cite{comparison} between the precessing mode
and the instability, this work is to be extended to treat 
finite-temperature systems~\cite{tomoyaaxisym1,sami}.

Concerning the splitting of the doubly quantized vortex,
the existence of two core-localized excitations
is confirmed throughout the calculated range of $L$.
The vortex pair always has an instability~\cite{pu,skryabin}
because systems with a vortex pair always support
core excitations with complex excitation
energies, \textit{c.f.} Fig.~\ref{fig:di_quadru_2}(d).

Throughout the analyses, the system displays two quadrupole excitations.
The excitation energies of the quadrupole modes
Figs.~\ref{fig:di_quadru_1}(b) and \ref{fig:di_quadru_2}(b)
are useful because these excitation
is used to measure the angular momenta
of the condensate experimentally.

\quad

%----------------------------------------------------------------
\section*{Acknowledgements}

Authors thank CSC Scientific Computing Ltd (Finland) for computer resources.
We are grateful to 
M. M\"ott\"onen, M. Nakahara, C. J. Pethick, T. P. Simula,
and S. M. M. Virtanen for stimulating discussions.
One of the authors (TI) is supported by 
the bilateral exchange programme between the Academy of Finland and
the Japan Society for the Promotion of Science.


\begin{thebibliography}{99}

\bibitem{pethick}
C. J. Pethick and H. Smith, \textit{Bose-Einstein Condensation in Dilute Gases}
(Cambridge University Press, Cambridge, England, 2002).


\bibitem{matthews}
M. R. Matthews, B. P. Anderson, P. C. Haljan, D. S. Hall,
C. E. Wieman, and E. A. Cornell,
Phys. Rev. Lett. \textbf{83}, 2498 (1999).

\bibitem{anderson}
B. P. Anderson, P. C. Haljan, C. E. Wieman, and E. A. Cornell,
Phys. Rev. Lett. \textbf{85}, 2857 (2000).

\bibitem{ens1st} 
K. W. Madison, F. Chevy, W. Wohlleben, and J. Dalibard, Phys. Rev. Lett. \textbf{84}, 806 (2000).

\bibitem{topological}
%Imprinting Vortices in a Bose-Einstein Condensate using Topological Phases
A. E. Leanhardt, A. G\"{o}rlitz, A. P. Chikkatur,
D. Kielpinski, Y. Shin, D. E. Pritchard, and W. Ketterle,
Phys. Rev. Lett. \textbf{89}, 190403 (2002)
% A. E. Leanhardt, A. G\"{o}rlits, A. P. Chikkatur, D. Kielpinski,
% Y. Shin, D. E. Pritchard, and W. Ketterle, e-print cond-mat/0206303.

\bibitem{nakahara}
M. Nakahara, T. Isoshima, K. Machida, S.-i. Ogawa, and T. Ohmi, Physica B \textbf{284}, 17 (2000);
T. Isoshima, M. Nakahara, T. Ohmi, and K. Machida, Phys. Rev. A \textbf{61}, 063610 (2000);
S.-i. Ogawa, M. M\"ott\"onen, M. Nakahara, T. Ohmi, and H. Shimada,
% e-print cond-mat/0202018
Phys. Rev. A \textbf{66}, 013617 (2002);
M. M\"ott\"onen, N. Matsumoto, M. Nakahara, and T. Ohmi,
J. Phys.: Cond. Matt. \textbf{14}, 29 (2002).

\bibitem{pu}
H. Pu, C. K. Law, J. H. Eberly, and N. P. Bigelow,
Phys. Rev. A \textbf{59}, 1533 (1999).
% Coherent disintegration and stability of vortices in trapped BC


\bibitem{dodd}
R. J. Dodd, K. Burnett, M. Edwards, and C. W. Clark,
Phys. Rev. A \textbf{56}, 587 (1997).

\bibitem{tomoyaaxisym1}
T. Isoshima, K. Machida, Phys. Rev. A \textbf{59}, 2203 (1999);

\bibitem{tomoyaaxisym2}
T. Isoshima, K. Machida, Phys. Rev. A \textbf{60}, 3313 (1999).

\bibitem{sami}
S. M. M. Virtanen, T. P. Simula, and M. M. Salomaa,
Phys. Rev. Lett. \textbf{86}, 2704 (2001).

\bibitem{sami:tpopov}
S. M. M. Virtanen, T. P. Simula, and M. M. Salomaa, 
Phys. Rev. Lett. {\bf 87}, 230403 (2001).
The time-dependent Hartree-Fock-Bogoliubov-Popov equations
reduces to time-dependent Bogoliubov equations at zero
temperature.

\bibitem{svidzinsky}
A. A. Svidzinsky and A. L. Fetter, \textbf{84}, 5919 (2000).

\bibitem{linn}
M. Linn and A. L. Fetter, Phys. Rev. A \textbf{61}, 063603 (2000).

\bibitem{feder}
D. L. Feder, A. A. Svidzinsky, A. L. Fetter, and C. W. Clark,
Phys. Rev. Lett. \textbf{86}, 564 (2001).
% Anomalous Modes Drive Vortex Dynamics in Confined BEC
% e-print cond-mat/0009086

\bibitem{comparison}
S. M. M. Virtanen, T. P. Simula, and M. M. Salomaa,
J. Phys: Condens. Matter \textbf{13}, L819 (2001).

\bibitem{butts}
D. A. Butts and D. S. Rokhsar, Nature \textbf{397}, 327 (1999).

\bibitem{guilleumas}
M. Guilleumas and R. Graham, Phys. Rev. A \textbf{64}, 033607 (2001).

\bibitem{chemical1}
The chemical potential $\mu$ is fixed for each of the procedures to obtain
the condensate wavefunction.
The procedure is repeated with 
slightly different value of $\mu$ until the result gives the
total particle number $N = 1\times 10^{10}\, \mathrm{m}^{-1}$.

\bibitem{chemical2}
The denominatar $\max(|\phi|)$ of Eq.~(\ref{eq:velocity}) does not vary much
[see peaks of densities in Figs.~\ref{fig:dns1}(a)]
because the trapping potential is fixed
and the chemical potential $\mu$ is chosen such that the
particle number $N$ becomes close to $N = 1\times 10^{10}\, \mathrm{m}^{-1}$.
So the denominator works as a normalization factor for the amplitude of the
condensate wavefunction.

\bibitem{fetter:vortex}
A. L. Fetter and A. A. Svidzinsky, e-print cond-mat/0102003.

\bibitem{vogels}
J. M. Vogels, K. Xu, C. Raman, J. R. Abo-Shaeer, and W. Ketterle,
Phys. Rev. Lett. 88, 060402 (2002).
% {\it Experimental observation of the Bogoliubov transformation
% for a Bose-Einstein condensed gas},
% e-print cond-mat/0109205

\bibitem{zambelli}
F. Zambelli and S. Stringari,
Phys. Rev. Lett. \textbf{81}, 1754 (1998).

\bibitem{chevy}
F. Chevy, K. W. Madison, and J. Dalibard, 
Phys. Rev. Lett. \textbf{85}, 2223 (2000).
%Measurement of the Angular Momentum of a Rotationg BEC


\bibitem{skryabin}
D. V. Skryabin, Phys. Rev. A \textbf{63}, 013602 (2000).


\end{thebibliography}
\end{document}